\documentclass{dimacs-l}
\usepackage{amssymb,amsmath,float} 
\usepackage{euscript}
\usepackage[dvips]{epsfig}

\newtheorem{theorem}{Theorem}
\newtheorem{lemma}[theorem]{Lemma}
\newtheorem{proposition}[theorem]{Proposition}

\newtheorem{definition}{Definition}

\renewenvironment{proof}[1]{\noindent {\it Proof~:} #1}
{\ \rule{1mm}{2mm}\medskip}

\newcommand{\remove}[1]{}

\newcommand\nd{\noindent}

\newcommand\nc\newcommand
\nc\bfa{{\mathbf a}}\nc\bfA{{\mathbf A}}\nc\cA{{\mathcal A}}
\nc\bfb{{\mathbf b}}\nc\bfB{{\mathbf B}}\nc\cB{{\mathcal B}}
\nc\bfc{{\mathbf c}}\nc\bfC{{\mathbf C}}\nc\cC{{\mathcal C}}
\nc\bfd{{\mathbf d}}\nc\bfD{{\mathbf D}}\nc\cD{{\mathcal D}}
\nc\bfe{{\mathbf e}}\nc\bfE{{\mathbf E}}\nc\cE{{\mathcal E}}
\nc\bff{{\mathbf f}}\nc\bfF{{\mathbf F}}\nc\cF{{\mathcal F}}
\nc\bfg{{\mathbf g}}\nc\bfG{{\mathbf G}}\nc\cG{{\mathcal G}}
\nc\bfh{{\mathbf h}}\nc\bfH{{\mathbf H}}\nc\cH{{\mathcal H}}
\nc\bfi{{\mathbf i}}\nc\bfI{{\mathbf I}}\nc\cI{{\mathcal I}}
\nc\bfj{{\mathbf j}}\nc\bfJ{{\mathbf J}}\nc\cJ{{\mathcal J}}
\nc\bfk{{\mathbf k}}\nc\bfK{{\mathbf K}}\nc\cK{{\mathcal K}}
\nc\bfl{{\mathbf l}}\nc\bfL{{\mathbf L}}\nc\cL{{\mathcal L}}
\nc\bfm{{\mathbf m}}\nc\bfM{{\mathbf M}}\nc\cM{{\mathcal M}}
\nc\bfn{{\mathbf n}}\nc\bfN{{\mathbf N}}\nc\cN{{\mathcal N}}
\nc\bfo{{\mathbf o}}\nc\bfO{{\mathbf O}}\nc\cO{{\mathcal O}}
\nc\bfp{{\mathbf p}}\nc\bfP{{\mathbf P}}\nc\cP{{\mathcal P}}
\nc\bfq{{\mathbf q}}\nc\bfQ{{\mathbf Q}}\nc\cQ{{\mathcal Q}}
\nc\bfr{{\mathbf r}}\nc\bfR{{\mathbf R}}\nc\cR{{\mathcal R}}
\nc\bfs{{\mathbf s}}\nc\bfS{{\mathbf S}}\nc\cS{{\mathcal S}}
\nc\bft{{\mathbf t}}\nc\bfT{{\mathbf T}}\nc\cT{{\mathcal T}}
\nc\bfu{{\mathbf u}}\nc\bfU{{\mathbf U}}\nc\cU{{\mathcal U}}
\nc\bfv{{\mathbf v}}\nc\bfV{{\mathbf V}}\nc\cV{{\mathcal V}}
\nc\bfw{{\mathbf w}}\nc\bfW{{\mathbf W}}\nc\cW{{\mathcal W}}
\nc\bfx{{\mathbf x}}\nc\bfX{{\mathbf Z}}\nc\cX{{\mathcal X}}
\nc\bfy{{\mathbf y}}\nc\bfY{{\mathbf Y}}\nc\cY{{\mathcal Y}}
\nc\bfz{{\mathbf z}}\nc\bfZ{{\mathbf Z}}\nc\cZ{{\mathcal Z}}
\nc\od{{\bar d}}\nc\ow{{\bar w}}\nc\odelta{{\bar\delta}}
\nc\ox{{\bar x}}\nc\oy{{\bar y}}\nc\ou{{\bar u}}

\newcommand\ff{{\mathbb F}}

\nc\dgv{\delta_{\text{\rm GV}}}
\nc\rcrit{R_{\text{\rm crit}}}
\nc\Esp{E_{\text{\rm sp}}}
\renewcommand\epsilon{\varepsilon}

\newcommand{\beeq}{\begin{eqnarray*}}
\newcommand{\eneq}{\end{eqnarray*}}
\newcommand{\eqdef}{\mbox{$\stackrel{\text{def}}{=}$}}

\newcommand{\x}{\mathbf x}
\newcommand{\y}{\mathbf y}
\newcommand{\e}{\mathbf e}
\renewcommand{\c}{\mathbf c}

\newcommand{\bigo}{\EuScript O}
\newcommand{\aux}{{\rm aux}}

\begin{document}
\title[Multilevel expander codes]
{Multilevel expander codes}

\author[A. Barg]{Alexander Barg$^\ast$}
\thanks{$^\ast$  Supported in part by NSF grant CCR 0310961.}
\address{Dept. of ECE, University of Maryland, College Park, MD 20742}
\email{abarg@ieee.org}
\author[G. Z\'emor]{Gilles Z\'emor}
\address{\'Ecole Nationale Sup\'erieure des 
T\'el\'ecommunications, 46 rue Barrault,
75 634 Paris 13, France} \email {zemor@enst.fr}

\copyrightinfo{2005}{American Mathematical Society}
\begin{abstract}
We define multilevel codes on bipartite graphs which have properties
analogous to multilevel serial concatenations. 
A decoding algorithm is described that corrects a proportion of
errors equal to half the Blokh-Zyablov bound on the minimum distance. The error
probability of this algorithm has exponent similar to that
of serially concatenated multilevel codes.
\end{abstract}

\maketitle

\section{Introduction}
Codes on graphs are presently actively studied both from
the standpoint of their parameters and the convergence properties of
various decoding algorithms. The codes on graphs 
studied in the present work originate in Tanner's paper \cite{tan81}
which suggested to index code bits  with
the edges of the graph and impose local constraints on the bits
indexed by the edges incident to any given vertex.
Following~\cite{tan81}, the local constraints are given by a set
of parity-check equations of a small error-correcting code which may be
different for every vertex of the graph.

In \cite{sip96}, codes on graphs were shown to correct
a proportion of errors growing linearly with block length 
under a simple iterative
decoding algorithm whose convergence region is related to the
expansion of the graph. Following \cite{sip96},
the term ``expander codes'' was adopted for this class of codes.
The line of work started by~\cite{sip96} was continued in several
directions. Studying codes on bipartite graphs,~\cite{zem01} 
introduced a new linear-time
iterative decoding
algorithm of expander codes which provided a better estimate of the fraction
of errors correctable by expander codes and linked these codes with other 
concatenated code constructions (this link was furthered in \cite{bar03d}).
Paper \cite{bar02c} showed that expander codes reach capacity of
the binary symmetric channel under the iterative decoding
algorithm of \cite{zem01}. The same paper also put forward the idea
of using two different codes as local constraints for the two parts
of the graph, which led to the construction of a family of expander
codes that match the performance of serially concatenated codes in the
sense of Forney~\cite{for66}. Independently, closely related results
were obtained in \cite{gur02}.
Code families with improved parameters and improved estimates of the error 
probability under iterative decoding were also studied in 
\cite{bar02c,bar04d,bil04,jan03,ska04}.

The purpose of the present paper is to extend the expander code construction
to the context of multilevel concatenations in the sense of Blokh and
Zyablov \cite{blo82}. This generalization pursues several goals. 
First, the asymptotic relative distance of the expander code family 
defined below is shown
to match the so-called Blokh-Zyablov bound $\delta_{BZ}(R)$ 
\cite{blo82}, which represents
a substantial improvement over the distance-rate tradeoff of single-level
concatenations \cite{zya71,bar02c}. An extension of the decoding algorithm
of \cite{bar02c} enables one to correct a fraction of errors that is
arbitrarily close to $(1/2)\delta_{BZ}(R)$. The same algorithm
guarantees a decrease of the error probability of decoding which approaches
the error exponent of multilevel concatenations of \cite{blo82}.
The main idea behind the construction of \cite{blo82} is the use
of a tower of nested codes in the inner level of concatenation which
simultaneously reach the typical performance of random linear codes.
Accommodating this idea to the bipartite-graph construction and setting
up a proper multilevel decoding algorithm are performed in a way
different from the standard approach.

The plan of the paper is as follows. Sect.\,\ref{sect:notation} is devoted
to some basic notation. Relevant background from multilevel serial
concatenations is reviewed in Sect. \ref{sect:serial}. 
A construction of multilevel parallel concatenations (multilevel 
expander codes) is introduced in Sect. \ref{sect:parallel}. 
The relative distance of the codes asymptotically approaches the 
Blokh-Zyablov bound. A multistage decoding algorithm of the codes 
defined suggested in Sect. \ref{sect:decoding} corrects a proportion
of errors that approaches $\delta_{BZ}(R)/2$ in time $O(N),$ where
$N$ is the code length. The error exponent of the algorithm approaches
the error exponent attainable by the multilevel serial concatenations 
of~\cite{blo82}.

\section{Notation}\label{sect:notation}
We assume transmission over a binary symmetric channel with transition
probability $p$, denoted below by BSC$(p)$. Let 
$h(x)=-x \log x-(1-x)\log(1-x)$ denote the binary entropy function
(the base of the logarithms is $2$ throughout).
Let $\dgv(R)=h^{-1}(1-R)$ denote the Gilbert-Varshamov (GV) relative 
distance for the rate $R$.

Below we will consider probabilities of different events expressed in the
form $\exp(-N E)$ where $N$ is a large positive number and $E$ a nonnegative
function of the code parameters. In such situations $E$ will be called
the {\em exponent} of the probability or the error exponent if the event we
have in mind is a decoding error.

Let $E_0(R,p)$ be the ``random coding exponent'' \cite{gal63}. For
$\rcrit \le R\le \cC$ we have
\(
E_0(R,p)=\Esp(R,p):=D(\delta_{\text GV}(R)\|p),
\)
where
\[
D(x\|y):=x\log (x/y)+(1-x)\log((1-x)/(1-y)),
\]
$\rcrit=1-h(\rho_0), \rho_0=\sqrt p/(\sqrt p+\sqrt {1-p}))$ is the
critical rate and $ \cC=1-h(p)$ is the channel capacity.
For rates $0\le R\le \rcrit$ the random coding exponent has the following
form:
\begin{eqnarray*}
E_0(R,p)&=&-\dgv(R)\log 2\sqrt{p(1-p)} \quad(0\le R\le R_x)\\
E_0(R,p)&=&D(\rho_0\|p)+\rcrit-R \quad(R_x\le R\le\rcrit),
\end{eqnarray*}
where $R_x=1-h(2\rho_0(1-\rho_0)).$
It is well known that for large code length, typical codes from the
random code ensemble
asymptotically achieve the GV bound and have error probability
of maximum likelihood decoding behaving as 
$$\exp(-N (E_0(R,p)-o(1))).$$

Let $q=2^t$ for some $t\ge 1.$
Given a binary code $A[n_0,k_0=2^t]$ and a $q$-ary code $B[n_1,k_1],$
their concatenation is a linear
code (mapping) defined as follows \cite{for66}:
    $$
      \ff_{2^{k_0}}^{k_1}\;\stackrel{B}{\longrightarrow}\; 
\ff_{2^{k_0}}^{n_1}\; \hookrightarrow\; (\ff_2^{k_0})^{n_1}
\;\stackrel{A}{\longrightarrow}\; (\ff_2^{n_0})^{n_1}.
    $$
We write $C=A \square B$ to denote the concatenation of codes $A$ and $B$
and note the parameters $[N=n_0n_1,K=k_0k_1]$ of the code $C$.

\section{Multilevel serial concatenations}\label{sect:serial}
We begin with recalling the construction and properties of serially
concatenated multilevel codes. These results are due to Blokh and
Zyablov \cite{blo82}. We review them here to emphasize analogies and
differences between multilevel serial and parallel concatenations. 
Then we proceed to a description of multilevel concatenations defined
on bipartite graphs, study their parameters and suggest a decoding
algorithm.


\subsection{Code construction}
A linear $m$th order concatenated code construction involves the following
ingredients: a binary $[n_0, k_0=n_0R_0,d_0]$ ``inner'' code $A$ and a set of 
$m$ ``outer'' codes $B_1,B_2,\dots,B_m,$ where the code $B_i, i=1,\dots,m$ 
is defined over an alphabet of size $q_i=2^{t_i}$. 
We assume that $k_0=\sum_{i=1}^m t_i$ and that each of the codes 
$B_i$ has parameters $[n_1,k_{1,i}=n_1 R_{1,i},d_{1,i}]$.

\begin{figure}[tH]
\epsfysize=6cm
\setlength{\unitlength}{1cm}
\begin{center}
\begin{picture}(5,7)
\put(-1,0){\epsffile{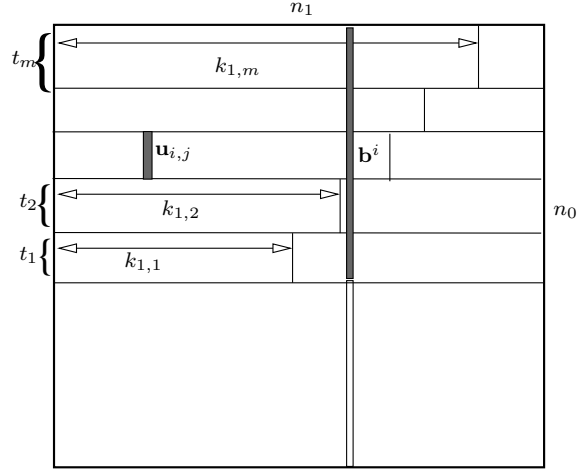}}
\put(0.7,4.2){{\footnotesize \mbox{$\bfu_{i,j}$}}}
\put(3.4,4.1){{\footnotesize \mbox{$\bfb^i$}}}
\put(0.3,2.7){{\footnotesize \mbox{$k_{1,1}$}}}
\put(0.8,3.4){{\footnotesize \mbox{$k_{1,2}$}}}
\put(1.5,5.3){{\footnotesize \mbox{$k_{1,m}$}}}
\put(2.5,6.1){{\footnotesize \mbox{$n_1$}}}
\put(6.0,3.4){{\footnotesize \mbox{$n_0$}}}
\put(-1.1,2.8){{\footnotesize \mbox{$t_1$}}}
\put(-1.1,3.5){{\footnotesize \mbox{$t_2$}}}
\put(-1.2,5.4){{\footnotesize \mbox{$t_m$}}}
\end{picture}
\end{center}
\caption{Construction of multilevel concatenations: codewords of $A$
   are read off verticaly, horizontal rectangles of width $t_i$ make up
   binary versions of the codes $B_i$}\label{fig:GCC}
\end{figure}

The length of the code $C$ equals $N=n_0n_1$ and its dimension is
\(
K=\sum_{i=1}^m k_{1,i}t_i.
\)
The code maps an information sequence $\bfu=(u_1,u_2,\dots,u_K)$
to the corresponding codeword by first mapping $\bfu$ to a vector in
$(\ff_{q_1})^{k_{1,1}}\times\dots\times(\ff_{q_m})^{k_{1,m}},$
then encoding it with the codes $B_i$, then
mapping the symbols of the obtained codewords back to binary digits,
and finally encoding them with the code $A$. The details are as follows:
let $K_0=0$ and for $i=1,\dots, m,$ let $K_i=\sum_{j=1}^{i}k_{1,j} t_j$.
Write the data vector $\bfu$ in the form 
$\bfu=(\bfu_1,\dots,\bfu_m),$ where 
$\bfu_i=(\bfu_{i,1},\dots,\bfu_{i,j},\dots,\bfu_{i,{k_{1,i}}})\in (\ff_2)^{k_{1,i}t_i}$ is a vector such that
    $$
     \bfu_{i,j}=(u_{K_{i-1}+(j-1)t_i+1},\dots,u_{K_{i-1}+jt_i})
    $$
(see Fig.~\ref{fig:GCC}). Next for $i=1,\dots, m,$ map $\bfu_i$ 
to a $k_{1,i}$-vector $\bfm_i$ over $\ff_{q_i}$ (this mapping is an
extension of the natural isomorphism) and encode the result with 
the code $B_i$. We obtain an $(m n_1)$-vector $\bfb\in
\ff_{q_1}^{n_1}\times\dots\times\ff_{q_m}^{n_1}$. 
For the second stage of
concatenated encoding, this vector is treated as a vector
in $F^{n_1},$ where $F=\ff_{q_1}\times\dots\times \ff_{q_m},$ 
and mapped onto $\ff_2$ using the isomorphism of additive groups
$F\cong \ff_2^{k_0}$. Denoting by $\langle \cdot\rangle_b$
the binary representation of a  vector, we can write
$\langle\bfb\rangle_b=\langle \bfb^1,\dots,\bfb^{n_1}\rangle_b,$
where $\langle\bfb^i\rangle_b, i=1,\dots, n_1$ is a binary $k_0$-vector.
Finally, each of these binary vectors is encoded with the code $A.$
The resulting vector $\bfc\in \ff_2^{n_0n_1}$
is a codeword of the $m$-th order concatenated code $C$.

The choice of the component codes and decoding of multilevel 
con\-ca\-te\-na\-tions
rely upon a decomposition of the code $A$ into a tower of nested
binary codes 
\begin{equation}\label{eq:tower}
(\ff_2)^{n_0}\supset A=A_{1}\supset A_{2}\supset\dots\supset
A_{m}\supset A_{m+1}=\{0\},
\end{equation}
where $A_{i}$ is an $[n_0,\sum_{j=i}^{m}t_i,d_{0,i}]$ binary linear code.
Let
     \begin{equation}\label{eq:matrix}
     \bfG_{0,1}=\left(\begin{array}{c}\bfG_m\\ \vdots \\ \bfG_1\end{array}
        \right)
     \end{equation}
be a generator matrix of $A$, where $\bfG_i, i=1,\dots, m$ is a binary
$(t_i\times n_0)$-matrix. Then the code $A_i, i=1,\dots, m$ 
is generated by
the matrix $\bfG_{0,i}=(\bfG_m^T\dots,\bfG_{i}^T)^T.$

Denote by $k_{0,i}=\sum_{j=i}^mt_j$ the dimension of the code
$A_i$ and by $R_{0,i}$ its rate.
Note that the rate of the code $C$ equals
    \begin{equation}\label{eq:ratei}
R=\sum_{i=1}^m(R_{0,i}-R_{0,i+1})R_{1,i} 
\qquad(R_{0,m+1}:=0).
     \end{equation}
Note also that if all the outer alphabets $q_i$ are the same and
equal to $q=2^t$, i.e., $t_i=t$ for all $i$, then the code $A_i$ has 
dimension $t(m-i+1)$ and the rate of the code $C$ can be written as
     \begin{equation}\label{eq:rate}
R=
\frac{R_0}m \sum_{i=1}^m  R_{1,i}.
     \end{equation}
The distance of the code $C$ is bounded below by $d(C)\ge \min_{1\le i\le m}
     d_{0,i}d_{1,i}.$

A view of multilevel concatenations as {\em direct sums} of single-level
concatenated codes is particularly helpful for a multilevel generalization
of expander codes. Let us introduce codes 
$A^{(i)}\cong A_i/A_{i+1}, i=1,\dots, m,$ where
$A^{(i)}$ is generated by $\bfG_{i}.$ Then the code $A$ can be represented
as a direct sum
    \begin{equation}\label{eq:direct-sum}
A=A^{(1)}\oplus A^{(2)}\oplus\dots\oplus A^{(m)}.
    \end{equation}
   We have the following theorem (see, e.g., \cite{dum98}).
\begin{theorem} $C={\bigoplus}_{i=1}^m A^{(i)}\square B_i$.
\end{theorem}

By definition of the codes $A^{(i)},$ 
the direct-sum decomposition of the code $A$ enables
us to write explicitly the cosets in the quotient spaces $A_i/A_{i+1},
i=1,\dots,m-1.$ Namely, let $\bfa=(\bfa_m,\dots,\bfa_1)$ be a message
vector of the code $A_1,$ where $\bfa_i \in (\ff_2)^{t_i}$ is the 
corresponding message vector of the code $A^{(i)}, i=1,\dots,m.$
Then the codeword $\bfx=\bfa \bfG_{0,1}$ can be written as
    \begin{equation}\label{eq:coset1}
       \bfx
         =(\bfa_m,\dots,\bfa_{2})\bfG_{0,2}+\bfa_1 \bfG_1.
    \end{equation}
This representation is useful in decoding of multilevel concatenations.

\medskip
\nd\begin{minipage}{\linewidth}
\nd{\sc Roster of component codes.}

\smallskip
1.The first-level (inner) binary codes
         $$
      A_i[n_0,k_{0,i}=\sum_{j=i}^m t_j,d_{0,i}],\quad i=1,2,\dots,m,
         $$

2. The second-level (outer) $q_i$-ary codes, $q_i=2^{t_i}$,
         $$
       B_i[n_1,k_{1,i},d_{1,i}],\quad i=1,2,\dots,m,
         $$

3. Binary codes
         $$
          A^{(i)}[n_0,k_0^{(i)}=t_i], \quad i=1,2,\dots,m.
         $$
These codes are not used directly in the code construction. They are helpful
in analyzing the decoding.
\end{minipage}
\medskip

We still have to specify which codes are used as components in this
construction. This choice is related to the complexity restrictions.
If the overall objective is polynomial construction and decoding complexity,
then the codes are chosen as follows. 
We assume that each of the codes $A_{i}, i=1,\dots, m$ attains the 
GV bound, i.e., that $R_{0,i}\ge 1-h(\delta_{0,i})-\epsilon$, 
and that its error exponent under maximum likelihood decoding approaches
the random coding exponent $E_0(R_{0,i},p).$ That such a tower of codes
exists is established by a standard technique, see \cite{blo82,dum98}.
The codes $B_1,\dots,B_m$ are taken to be some algebraic $q$-ary codes 
that afford a decoding algorithm of complexity polynomial
in the length $n_1,$ typically Reed-Solomon or algebraic geometry codes.

\subsection{Decoding}
Decoding proceeds in $m$ stages. 
Every stage consists of two steps, namely, applying maximum likelihood
decoding to the code $A_{j}$ and then decoding  $B_j$ by an algebraic
procedure.

Let $\bfy=\bfy_1$ be the vector received from the channel. We will write
$\bfy_1=(\bfy_{1,1},\dots,\bfy_{1,n_1}),$ where for every $j$ the
vector $\bfy_{1,j}\in (\ff_2)^{n_0}.$
In stage one, every vector $\bfy_{1,j}, j=1,2,\dots, n_1$ is decoded 
with the code $A_{1},$
employing one of the possible decoding procedures for this code.
Next, for every $j,$ $t_1$ information bits are extracted from the 
decoded message sequence 
(we will assume that the decoding algorithm of the codes $A_{i}$
always outputs a codeword; if not, the procedure is easily modified to 
handle erasures). The $n_1$ groups of these message bits,
viewed as an $n_1$-vector over $F_{2^{t_1}}$ constitute
a ``received'' word of the code $B_1.$ This word is decoded with the code
$B_1$; denote the outcome of this decoding by 
$\tilde\bfb_1=(\tilde b_{1,1},\dots,\tilde b_{1,{n_1}}).$
Note that if $\tilde\bfb_1\ne\bfb_1,$ the overall decoding procedure has
ended in error; this will be a part of the error event.

Stage two (and every subsequence stage) is not much different from stage
one; however we have now to decode the pair of codes $A_{2}, B_2.$
This requires a transformation of the received word $\bfy_1$ to a vector
$\bfy_2=(\bfy_{2,1},\dots,\bfy_{2,n_1})$ which relates to the code
$A_{2}$ in the same way as $\bfy_1$ relates to $A_{1}.$ We compute
     \begin{equation}\label{eq:coset}
\bfy_{2,j}=\bfy_{1,j}+\bfz_{1,j},
     \end{equation}
where $\bfz_{1,j}=\langle\tilde b_{1,j}\rangle_b \bfG_1$, and where 
$\langle\tilde b_{1,j}\rangle_b\in (\ff_2)^{t_m}$ 
is the binary representation of the symbol $b_{1,j}.$
This transformation is easily understood in the absence of errors:
in that case this is simply the coset decomposition of the nested codes 
of (\ref{eq:coset1}).
The rest of this decoding stage is the same as in stage one, with
$(A_{1},B_1)$ replaced with $(A_{2},B_2).$

\subsection{Error probability and the choice of parameters.}
While we have chosen the component codes in the construction of $C$,
we still have some freedom in choosing the parameters of these codes.
This choice is optimized relying on performance estimates of
decoding. Let us focus on the error probability.
Decoding the $i$th level of the code $C$ is not very different from
decoding a standard (one-level) concatenated code. We assume maximum
likelihood decoding of the code $A_{i}$ and Generalized Minimum
Distance decoding of the code $B_i$ \cite{for66}. The error exponent
upon decoding of the code $A_{i}$ can be assumed to approach
$E_0(R_{0,i},p)$, and the overall error exponent obtained upon GMD decoding
of the code $B_i$ then equals $E_i=E_0(R_{0,i},p)(1-R_{1,i}).$ 
The overall error exponent for the 
code $C$ then equals
\begin{equation}\label{eq:bb}
E^{(m)}(R,p)=\min_{1\le i\le m} (E_0(R_{0,i},p)(1-R_{1,i})).
\end{equation}
which can be further optimized on the choice of the rates $R_{0,i},
R_{1,i}$
given that they must satsify relation (\ref{eq:ratei}).

Let us assume that all the $t_i$ are equal and denote their common value
by $t$, in which case
\(
R_{0,i}=t(m-i+1)/n_0.
\)
The following results \cite{blo82} are derived under the assumption that
all the terms under the minimum in (\ref{eq:bb}) are the same and 
equal to $E^{(m)}(R,p).$ 
We then have the following proposition whose proof is included for later use.
\begin{proposition}\label{prop:m-exp}\cite{blo82}
\begin{equation}\label{eq:m-exp}
E^{(m)}(R,p)=\max_{R\le R_0\le \cC} \frac{R_0-R}{\frac{R_0}{m}
\sum_{i=1}^m(E_0(\frac imR_0,p))^{-1}}.
\end{equation}
\end{proposition}
\begin{proof}
{From} (\ref{eq:bb}) and the assumptions above,
\[
R_{1,i}=1-\frac{E^{(m)}(R,p)}{E_0(R_{0,i},p)} \quad(i=1,\dots,m);
\]
together with (\ref{eq:rate}) this gives
\[
Rm= R_0\Big[ m-E^{(m)}(R,p)\sum_{i=1}^m (E_0(R_{0,i},p))^{-1}\Big].
\]
Solving for $E^{(m)}$ and using the expression for $R_{0,i}$, 
we obtain the claim.
\end{proof}

For $m=1$ this bound turns into the so-called Forney bound \cite{for66}
      \begin{equation}\label{eq:fex}
        E^{(1)}(R,p)=\max_{R\le R_0\le \cC} E_0(R_0,p)(1-R/R_0).
      \end{equation}
For all rates $R\in (0,1),$ increasing the order of concatenation
improves the bound on the error exponent, namely
    $$
        E^{(m)}(R,p)< E^{(m+1)}(R,p), \qquad m=1,2,\dots .
    $$

Letting $m\to\infty$ in (\ref{eq:m-exp}), we obtain the error exponent
of concatented codes of infinite order:
   \begin{equation}\label{eq:inf-exp}
     E^{(\infty)}(R,p)=\max_{R\le R_0\le \cC} (R_0-R)
          \Big[\int_0^{R_0} \frac{dx}{E_0(x,p)}\Big]^{-1}.
   \end{equation}
This is the {\em Blokh-Zyablov exponent} \cite{blo82}. Note an alternative,
parametric expression for it, obtained upon performing the maximization:
\begin{align*}
E^{(\infty)}(R,p)&=E_0(\alpha,p)\\
R&= \alpha-E_0(\alpha,p)\int_0^{\alpha} \frac{dx}{E_0(x,p)} 
\qquad(0\le \alpha\le 1-h(p)).
\nonumber
\end{align*}
The convergence of $E^{(m)}$ to $E^{(\infty)}$ is uniform for
$R\in [0,\cC-\epsilon].$ In the neighborhood of capacity the function
under the intergal in (\ref{eq:inf-exp}) has a singularity: a little
analysis shows that for $E^{(m)}$ to approximate well $E^{(\infty)}$,
the order $m$ has to grow faster than $1/\epsilon.$


\subsection{Minimum distance}
The minimum (relative) distance of an $m$-level code for 
$N\to\infty$ approaches the bound \cite{blo82}
    \begin{equation}
       \delta^{(m)}(R)=\max_{R\le R_0\le 1}
           \frac{m(R_0-R)}{R_0\sum_{i=1}^m \big(\dgv\big(\frac{i}m R_0
               \big)\big)^{-1} }.
    \end{equation}
For $m=1$ this expression turns into the so-called {\em Zyablov bound} 
\cite{zya71}
    \begin{equation}\label{eq:zb}
      \delta_{\text Z}(R)=\max_{R\le R_0\le 1} \dgv(R_0)(1-R/R_0),
    \end{equation}
and for $m\to\infty$ it becomes the {\em Blokh-Zyablov bound}, which 
is easier to write expressing the rate $R$ as a function
of the relative distance. We obtain
\begin{equation}\label{eq:bz}
R_{\text{BZ}}(\delta)= 1-h(\delta)-\delta \int_0^{1-h(\delta)}
   \frac{dx}{\dgv(x)}.
\end{equation}
\remove{
Note that for $\epsilon \to 0+$ we have the following asymptotic results:
\begin{alignat*}{2}
\delta&=\epsilon & R_{\text{BZ}}&=1-\frac{\ln 2}2\,\epsilon(\log_2 \epsilon)^2 \\
\delta&=\frac12-\epsilon  &\qquad R_{\text{BZ}}&=(4/3\ln2)\epsilon^3.
\end{alignat*}
Compare this with random codes that meet the GV bound: we then get 
$R_{\text{GV}}=1+\epsilon\log_2\epsilon$ for small $\delta$ and
$R_{\text{GV}}=(2/\ln 2)\epsilon^2$ for large $\delta.$}

The Blokh-Zyablov bound can be improved by using long
algebraic geometry codes as the outer codes $B_1, \dots, B_m.$ 
The results have been computed  only for $m=1$ \cite{kat84a}. 
\remove{
It is notable 
that under polynomial-time decoding, and for $\delta=(1/2)-\epsilon$
the rate of the codes in \cite{kat84a} behaves
as $1+2\epsilon\log_2\epsilon,$ i.e., practically 
matches the rate of random codes. For small $\delta$ the results of 
\cite{kat84a} are inferior to the Blokh-Zyablov bound.}

\subsection{Complexity of multilevel concatenated codes}
The decoding complexity of one-level concatenated codes that meet
the Forney bound is $O(n^2).$ The same complexity estimate is valid
for decoding of one-level concatenations that correct the proportion
of errors that asymptotically approaches $\delta_{\text Z}$.
Turning to the complexity of $m$-level concatenations with outer 
Reed-Solomon codes, we set $n_1=2^{R_{0,1}n_0/m}$. Let $m$ grow as
$\log n_0$ and consider codes of relative distance $\delta$ and rate
$R,$ where $R$ and $\delta$ are related by (\ref{eq:bz}). 
Their decoding complexity for correcting
a $\frac12 \delta_{\text BZ}$ proportion of errors, where
$\delta_{\text{BZ}}$
is the solution of (\ref{eq:bz}) with respect to $\delta,$ or for
achieving the error exponent $E^{(\infty)}$ can be bounded above
as $n^{1+\log\log n/(1-h(\delta))}$.

\section{Parallel concatenations}\label{sect:parallel}

\subsection{Single-level constructions}
In its basic version, a bipartite-graph (BG) code is defined as follows.
Let $G(V=V_0\cup V_1,E)$ be a $\Delta$-regular bipartite graph with
$|V_0|=|V_1|=n.$ Let $A[\Delta, R_0\Delta], B[\Delta,R_1\Delta]$
be additive binary or $q$-ary codes. 
Let us fix an ordering of the edges of $E$
and construct a bipartite-graph code $C(G;A,B)$ of length $N:=|E|=n\Delta$
whose coordinates are in a one-to-one correspondence with the edges in $E$.
For a vector $\bfx\in C$ denote by $\bfx_v$ a projection of $\bfx$
on the edges incident to a vertex $v\in V$: in other words, if $E(v)$
denotes the set of edges incident to $v$, then $\bfx_v=(x_j)_{j\in E(v)}$.

A vector $\bfx$ is a code vector of $C$ if 
\begin{enumerate}
\item for every $v\in V_0$, the vector $\bfx_v\in A,$
\item for every $w\in V_1$, the vector $\bfx_w\in B$.
\end{enumerate}

The basic iterative procedure used to decode expander codes 
was introduced in \cite{zem01} and further studied in
\cite{bar02c,ska04}. It is described as follows.

\medskip

\noindent
{\bf Basic decoding scheme.}
Given a vector $\bfy\in \{0,1\}^N,$
a left decoding round $L$ consists of decoding in parallel with the code $A$
the subvectors $\bfy_v$ for all $v\in V_0$. Likewise, a right decoding
round $R$ applies decoding with the code $B$ to the subvectors $\bfy_w$
for every $w\in V_1.$ Decoding of the component codes is assumed maximum
likelihood. The {\em basic expander decoding} scheme consists of performing
successive decoding steps of the form $\bfy_{i+1}=R(L(\bfy_i)), i=0,1,\dots.$
Decoding terminates by either encountering a fixed point or performing
$O(\log n)$ decoding steps.

\medskip

In each level of the multilevel construction we will use a modified 
definition of bipartite-graph codes. Introducing this modification
enables one to match the performance of serial concatenations in 
the parallel case both for single-level concatenations \cite{bar03d} 
and multilevel codes as described below.
Let $G(V,E)$ be a bipartite graph whose parts are $V_0$ (the left
vertices) and $V_1\cup V_2$ (the right vertices), where
$|V_i|=n$ for $i=0,1,2.$ We will choose both subgraphs 
$G_i=(V_0\cup V_i, E_i), i=1,2$  to be
   regular, of degrees $\Delta_1$
and $\Delta_2$ respectively.  Thus, the degree of the left vertices 
is $\Delta,$ the degree of the vertices in $V_1$ is $\Delta_1,$ and
   the degree of vertices in $V_2$ is $\Delta_2-\Delta_1.$ 
Let $\lambda$ be the second
largest eigenvalue of the subgraph $G_1$ spanned by the vertex sets
$V_0$ and $V_1$. We will assume that $\lambda$ is small compared to $\Delta_1$,
for instance, that $\lambda=O(\sqrt{\Delta_1}).$

Let us fix an arbitrary
ordering of the edges in $E$. Note that $|E|=n\Delta.$ We are going
to construct a linear code $C$ of length $N=n\Delta t$, where $t$
is an integer constant, in such a way that every edge in $E$ corresponds
to $t$ coordinates of $C$.

For a given vertex $v \in V_0$ 
we denote by $E(v)$ the set of all
edges incident to it and by $E_i(v)\subset E(v), i=1,2$ the subset of 
edges  of the form $(v,w)$, where $w\in V_i$. 
The ordering of the edges on
$v$ defines an ordering on $E_i(v).$

Let $A$ be a $[t\Delta, R_0t\Delta,d_0=t\Delta \delta_0]$ linear binary code 
of rate $R_0=\Delta_1/\Delta$. The code $A$ can be also viewed as a $q$-ary
additive $[\Delta, R_0\Delta]$ code, $q=2^t$. Let $B$ be a $q$-ary 
$[\Delta_1, R_1\Delta_1,d_1=\Delta_1\delta_1]$ additive code. 
We will also need an auxiliary $q$-ary code $A_{\text{aux}}$
of length $\Delta_1$. Every edge of the graph will be associated with
$t$ bits of the codeword of the code $C$ of length $N=nt\Delta.$
The code $C$ is defined as the set of vectors
$\bfx=\{x_1,\dots, x_N\}$ such that

\begin{enumerate}
\item For every vertex $v\in V_0$ the subvector $(x_j)_{j\in E(v)}$ is
a ($q$-ary) codeword of $A$ and the set of coordinates $E_1(v)$ is an
information set for the code $A$;
\item For every vertex $v\in V_1$ the subvector $(x_j)_{j\in E(v)}$
is a codeword of $B$;
\item For every vertex $v\in V_0$ the subvector $(x_j)_{j\in E_1(v)}$
is a codeword of $A_{\text{aux}}.$
\end{enumerate}

This code family was introduced in \cite{bar04a} and studied extensively
in \cite{bar03d,bar04d}. In particular, \cite{bar03d} introduced a
modified iterative decoding algorithm of the code $C$ that uses expansion 
properties of the graph $G_1$ together with 
passing reliability information gathered
from decoding of the left codes to the right code
decoders. More precisely, the algorithm is described as follows.

\medskip

\paragraph{\bf Modified decoding scheme.} Let $\bfy\in\{0,1\}^N$ be the vector 
received from the channel.
In the first step, for every vertex $v\in V_0$, the left decoder
computes, for every neighboring vertex $w\in V_1$ and for every
$q$-ary symbol $b$, the quantity $d_{\{v,w\}}(b)$ which is the minimum
distance of $\bfy_v$ to a codeword of $A$ with symbol $b$ in
coordinate $\{v,w\}$. Then this quantity is passed on through edge
   $\{v,w\}$ to the right decoder at vertex $w$.

In the second step, for every right vertex $w$, the right decoder
finds the codeword $\bfc =(c_j)_{j\in E(w)}$ of $B$ that minimizes
$\sum_{j\in E(w)}d_j(c_j)$ (this is an iteration of min-sum decoding).
The right decoder then writes the symbols of $\bfc$ on its edge set
$E(w)$.

The decoder then reverts to the basic iterative procedure applied to
the basic bipartite-graph code $C( (V_0\cup V_1,E_1); A_{\text{aux}}, B)$,
using the decoding results of the second step as its starting values.
If this procedure succeeds, then an information set of $A$ is
recovered at every left vertex, and the whole original codeword $\bfx$
can be rederived.

\medskip

The complexity of the
algorithm is $\bigo(N)$, similarly to the basic expander decoding 
scheme.
The properties of this algorithm together with the parameters
of the code family are summarized as follows.
\begin{theorem}\label{thm:sl} \cite{bar03d} 
The code $C$ has the parameters $[N=nt\Delta,RN,D],$
where $R\ge R_0R_1-R_0(1-R_{\text{aux}})$ and
      $$
        D\ge \delta_0\delta_1 \left(1-\frac{\lambda}{d_\aux}\right)
               \left(1-\frac{\lambda}{2d_1}\right)N.
      $$
Let $n\to \infty$ and let $R$ be fixed. 
For any $\epsilon>0$ there exists sufficiently large
but constant values of $\Delta$ and $t$ such that 
$D/N\geq \delta(R)-\epsilon$, 
where $\delta(R)$ is the Zyablov bound {\rm(\ref{eq:zb})}. 
The decoding algorithm of {\cite{bar03d}} has error exponent given by
{\rm(\ref{eq:fex})}. The algorithm corrects a proportion of errors that 
approaches
$\delta_{\rm Z}(R)/2.$
\end{theorem}
Correcting a fraction $\delta_{\rm Z}(R)/2$ of errors in linear time 
was independently obtained in \cite{gur02}.
An alternative view of the above code construction 
was suggested in \cite{ska04}. 
The approach of \cite{ska04} also made it possible to use
Generalized Minimim Distance decoding of \cite{for66}
in the iterative expander decoding procedure. Thereby \cite{ska04} 
obtained a different proof of the error correction radius
and the error exponent for the above code construction.

\subsection{A multilevel construction}

Let $G$ be a bipartite graph with the vertex set $V$ and the edge set $E$.
The sets $V$ and $E$ are partitioned respectively
as 
\beeq
    V & = & V_0\cup V_1\cup\ldots\cup V_m\cup V_{m+1}\\
    E & = & E_1\cup\ldots\cup E_m\cup E_{m+1}
\eneq
where $E_i$ is the set of edges between $V_0$ and $V_i$, $i=1,\ldots, m+1$. 
The cardinalities of the sets $V_i$, $i=0,\ldots, m+1$, are all
taken to be the same, $|V_0|=|V_1|=\ldots =|V_{m+1}|=n$, and each edge set
$E_i,\,i\geq 1$,
is taken to define a regular bipartite expander graph $G_i$ on $V_0\cup V_i$
of degree $\Delta_i$. 
Every vertex $v\in V_0$ has therefore degree 
$\Delta\eqdef \Delta_0 \eqdef \Delta_1+\cdots \Delta_m+\Delta_{m+1}$ 
in the resulting graph, and the total number
of edges equals $|E|=n\Delta$. We will assume that each edge in $E$
carries $t$ bits of the codeword of the multilevel BG code $C$
for some positive constant $t$, so the code length is $N=tn\Delta.$

As above, let us fix an arbitrary order of the edges in $E.$
The edges $E(v)$ adjacent to a vertex $v\in V_0$ are partitioned 
into disjoint subsets as follows
      $$
         E(v)=\bigcup\limits_{i=1}^{m+1} (E(v) \cap E_i).
      $$
To define the code $C$, we need several component codes. 
Let $A$ be a binary linear code of length $t\Delta$ and dimension $R_0t\Delta$.
Referring to the representation of the graph $G$ shown in Figure 1
below, we call this code the ``left'' code.
The code $A$ can be also viewed as a $q$-ary additive code
of length $\Delta$ and dimension $R_0\Delta,$ where $q=2^t.$

Suppose again that $A$ affords a nested decomposition
     \begin{equation}\label{eq:subcodes2}
     A=A_{1}\supset A_{2}\supset
     \cdots \supset A_{m}\supset A_{m+1}=\{0\},
    \end{equation}
where this time the codes $A_i$ are viewed as binary or $q$-ary depending
on the context. The rate of the code $A_i$ (binary or $q$-ary) is 
$R_{0,i}=\sum_{j=i}^m \Delta_j/\Delta.$
The decomposition (\ref{eq:subcodes2}) 
is chosen to have properties similar to those of (\ref{eq:tower});
in particular, each code $A_i$ is assumed to approach the GV bound on
the relative distance. The binary version of the code $A_i$ is assumed
to have error exponent close
to the random coding exponent $E_0(R_{0,i},p)$ under maximum likelihood
decoding on the BSC.

As above, let $A$ be decomposed into a direct sum
    $$
    A=A^{(1)}\oplus A^{(2)}\oplus\cdots\oplus A^{(m)}
    $$
where each $A^{(i)}$ is a $[t\Delta,t\Delta_i]$ binary linear code. 
This decomposition is similar to the decomposition (\ref{eq:direct-sum}). 
The codes $A_i$ and $A^{(i)}$ can be also viewed as $q$-ary additive codes, 
with obvious adjustments to their
parameters.
Having the $q$-ary representation in mind,
denote by $\bfG_i$ a generator
matrix of the code $A^{(i)}, i=1,\dots, m.$ We will assume that these
matrices are chosen in some fixed way. 
The code $A_i$ is generated by 
$\bfG_{0,i}=(\bfG_i^T,\bfG_{i+1}^T\dots,\bfG_{m}^T)^T$ and therefore can be written
as a direct sum
     $$
      A_i=A^{(i)}\oplus A^{(i+1)}\oplus\dots\oplus A^{(m)}.
     $$
For any $\bfc\in A_i$, let $\bfc = \c^{(i)}\oplus \cdots \oplus 
\c^{(m)}$ be a direct sum decomposition of $\bfc$ and let
$\bfa^j\in \ff_q^{\Delta_j}, j=i,\dots, m$ 
be the $q$-ary message vector that corresponds to
$\bfc^{(j)},$ i.e., the unique vector such that 
$\bfa^j\cdot\bfG_{j} = \bfc^{(j)}$.

We will also need $m$ auxiliary $q$-ary codes $A_{i,\aux}$ of lengths
$\Delta_i$ and rates $R_{\aux,i},i=1,\dots, m.$ For each $i,$ the value
$R_{\aux,i}$
is assumed to be close to one, in particular, 
$R_{\aux,i}=1-\bigo(1/\sqrt{\Delta})$.

Finally, we need $m$ ``right'' $q$-ary codes $B_i[\Delta_i,R_{1,i}\Delta_i,d_{1,i}].$

\begin{figure}[tH]
\epsfysize=8.3cm
\setlength{\unitlength}{1cm}
\begin{center}
\begin{picture}(5,8)
\put(0,0){\epsffile{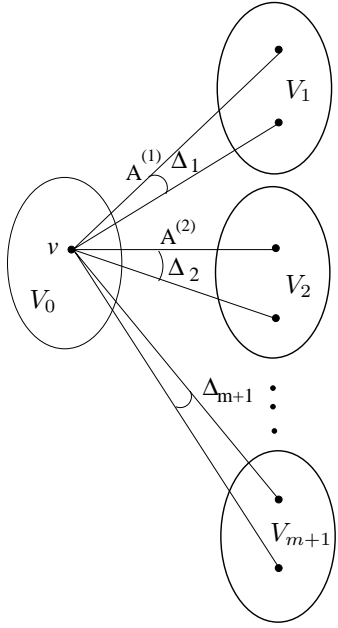}}
\put(0.3,4.2){$V_0$}
\put(3.7,7.0){$V_1$}
\put(3.7,4.4){$V_2$}
\put(3.5,1.1){$V_{m+1}$}
\end{picture}
\end{center}
\caption{Coding of $m$-level bipartite-graph codes}\label{fig:tree}
\end{figure}

\begin{definition} An $m$-level bipartite-graph code $C$ of length 
$N=nt\Delta$ is a set of vectors $\bfx\in (\ff_{2})^N$ that satisfies
the following conditions.\\[2mm]
$(1)$ For every $v\in V_0, \bfx_v\in A$.\\[1mm]
$(2)$ For every $v\in V_0$, let
$\bfx_v=\bfx_v^{(1)}\oplus\dots\oplus \bfx_v^{(m)}$ be the direct-sum
decomposition of the vector $\bfx_v.$ For $i=1,\dots, m$, let
the $q$-ary message vector $\bfa_v^i$ be defined by the equality
$\bfa_v^j\cdot\bfG_{j} = \bfx_v^{(j)}$. Then
$\bfa_v^i$ is a code vector of $A_{i,\aux}.$\\[1mm]
$(3)$
Let $\bfa$ be the $q$-ary vector of length 
$nR_0\Delta = n\sum_{i=1}^m\Delta_i$ deduced from $\bfx$ by writing,
for every $v\in V_0$, the vector $\bfa_v^i$ on the edge set 
$E(v)\cap E_i$. Then 
for every $i=1,\dots, m$ and every $w\in V_i,$ the vector 
$\bfa_w\in B_i.$
\end{definition}
The construction of multilevel bipartite-graph codes is illustrated in Fig.~\ref{fig:tree}.

\medskip
\nd\begin{minipage}{\linewidth}
\nd{\sc Roster of component codes.}

\smallskip
1. The ``left''  binary codes
         $$
      A_i[t\Delta, t k_{0,i}=t\sum_{j=i}^m \Delta_i,d_{0,i}],\quad i=1,2,\dots,m,
         $$
that form a tower of nested codes (\ref{eq:subcodes2}). 
These codes can be also viewed as additive codes over $\ff_{2^t}$
with parameters $[\Delta,k_{0,i}]$.

2. The ``right'' codes
         $$
       B_i[\Delta_i,k_{1,i}=R_{1,i}\Delta_i,d_{1,i}=\delta_{1,i}\Delta_i]
\quad\text{ over } \ff_{2^{t}}, i=1,2,\dots,m,
         $$

3. Binary codes
         $$
          A^{(i)}[t\Delta,t \Delta_{i}], \quad i=1,2,\dots,m
         $$
such that $A_i=\bigoplus\limits_{j=i}^{m} A^{(j)}.$
The rate of the code $A^{(i)}$ equals $R_0^{(i)}=R_{0,i}-R_{0,i+1},$
where $R_{0,m+1}:=0.$
These codes can be also viewed as additive codes over $\ff_{2^t}$
with the parameters $[\Delta,\Delta_{i}]$.

4. $q$-ary codes $A_{i,\aux}[\Delta_i,R_{i,\aux}\Delta_i],\;i=1,2,\dots,m.$
\medskip
\end{minipage}

\subsection{Parameters} 
The rate of the code $C$ is determined in the following
\begin{proposition} For any $\epsilon>0$ there exists a graph 
$G$ and a choice of component codes of the code $C$ such that its rate
satisfies
     $$
       R(C)\ge \sum_{i=1}^m(R_{0,i}-R_{0,i+1})R_{1,i}-\epsilon,
     $$
where $\epsilon=\bigo(1/\sqrt{\Delta})$.
\end{proposition}
\begin{proof} From the definition of the code $C$,
     $$
     tn\Delta(1-R(C))\le tn\Delta(1-R_0)+
         tn \sum_{i=1}^m\Delta_i(1-R_{i,\aux}) +tn
            \sum_{i=1}^m \Delta_i(1-R_{1,i}),
     $$
   from which,
    $$
      R\ge \sum_{i=1}^m(R_{0,i}-R_{0,i+1})R_{1,i}-R_0+\sum_{i=1}^m
          (R_{0,i}-R_{0,i+1})R_{i,\aux}.
    $$
Choosing the values of $\Delta_i$ and the auxiliary codes so that
$\min_i R_{i,\aux}\ge 1-\epsilon/R_0,$ we obtain the claim.
\end{proof}

Observe that the rate of multilevel parallel concatenations comes close
to the value of the rate of their serial counterparts (\ref{eq:ratei}).

The distance of the code $C$ is estimated from the distances of expander
codes supported by the graphs $G_i, i=1,\dots,m.$ Extending the proof
of Theorem \ref{thm:sl} to the $m$-level case, we obtain
\begin{theorem} For any $\epsilon>0$ there exist a graph $G$ and
a choice of the component codes of the multilevel construction
such that the distance $D$ of the code $C$ satisfies
     $$
       \frac DN\ge \min_{1\le i\le m} \delta_{0,i}\delta_{1,i}-\epsilon.
     $$
\end{theorem}

In particular, the results of this section imply 
that for $n\to\infty$
the family of multilevel parallel concatenations approaches
the Blokh-Zyablov bound (\ref{eq:bz}).
More precisely, given an $\epsilon>0$ and a value of the rate $R$, it is possible to find large
(but independent of $n$)
values of $\Delta,t,m$, a set of codes $A,B_i,i=1,\dots,m$, and
a family of graphs $G_{\{n\}}$ with $n$
vertices in each component such that as $n\to\infty,$ the relative distance
of the multilevel bipartite graph code will be within $\epsilon$ of
the quantity $\delta_{\text{BZ}}(R).$

\subsection{Decoding}\label{sect:decoding}
Let $\bfx$ be the transmitted codeword and let $\bfy=\bfy_1$ 
be the vector received from the channel. For every vertex $v\in V_0$,
let $\bfx_v=\bfx_v^{(1)}\oplus\dots\oplus \bfx_v^{(m)}$ be the direct
sum decomposition of $\bfx_v$.
The decoding proceeds in $m$ stages, the purpose of stage $i$ being
the recovery of the vector $\bfx_v^{(i)}$.

The first stage consists of three steps, similar to the three steps of
the modified decoding scheme of single-level constructions.

In the first step, given the vectors $\bfy_v, v\in V_0$, the decoder
computes 
for every neighboring vertex $w\in V_1$ and for every
$q$-ary symbol $b$, the quantity $d_{\{v,w\}}(b)$ which is the minimum
distance of $\bfy_v$ to a codeword
    $$\bfc_{1,v}=\bfc_{1,v}^{(1)}\oplus\dots\oplus \bfc_{1,v}^{(m)}$$
of $A$ {\em such that the vector $\bfa_v$ defined by
    $$\bfa_v\cdot\bfG_{1} = \bfc_v^{(1)}$$
has symbol $b$ in coordinate  $\{v,w\}$.}
Then this quantity is passed on along edge
   $\{v,w\}$ to the right decoder at vertex $w$.

In the second step, 
for every right vertex $w\in V_1$, the right decoder
finds the codeword $\hat\bfb_1 =(\hat b_{1,j})_{j\in E(w)}$ of the code
$B_1$ that minimizes
$\sum_{j\in E(w)}d_j(\hat b_{1,j})$ (this is an iteration of min-sum decoding).
The right decoder then writes the symbols of $\hat\bfb_1$ on its edge set
$E(w)$.

The decoder then reverts to the basic iterative decoding scheme applied to
the basic bipartite-graph code $C( (V_0\cup V_1,E_1); A_{1,aux}, B_1 )$.
If this procedure succeeds, then an information vector of $A^{(1)}$ is
recovered at every left vertex, i.e. for every $v\in V_0$, the decoder
has found $\bfa_v^1$ such that $\bfa_v^1\cdot\bfG_{1} = \bfx_v^{(1)}$.

Call $\hat \bfa_v^{1}$ the vector actually recovered at vertex $v$ at this
point of the decoding procedure. Next for every $v\in V_0$ we compute
        \begin{equation}\label{eq:coset2}
          \bfy_{2,v}=\bfy_v +\hat \bfa_v^{1}\bfG_{1}.
        \end{equation}

The vectors $\bfy_{2,v}$ form a vector $\bfy_2$ which is submitted to
the second stage of the decoding procedure. 
Note that if this first
stage of decoding was successful, then every vector $\bfy_2$ is such
that $\bfy_{2,v}$ is equal to the channel error vector added to
$$\bfx_v^{(2)}\oplus\dots\oplus \bfx_v^{(m)}.$$

The second (and every 
subsequent) stage is now the same as stage one except that
we use the code $A_2$ to decode $\bfy_2$. Let us elaborate.

In step one of stage $2$ (respectively $i$), the vectors $\bfy_{2,v}$ 
($\bfy_{i,v}$) compute, for every neighboring vertex $w\in V_2$ ($w\in
V_i$) and for every $q$-ary symbol $b$ the quantity $d_{\{v,w\}}^2(b)$
($d_{\{v,w\}}^i(b)$) which is the minimum distance of $\bfy_{2,v}$
($\bfy_{i,v}$) to a codeword
    $$\bfc_{2,v}=\bfc_{2,v}^{(2)}\oplus\dots\oplus \bfc_{2,v}^{(m)}
\hspace{1cm}(\bfc_{i,v}=\bfc_{i,v}^{(i)}\oplus\dots\oplus \bfc_{i,v}^{(m)}) $$
of $A_2$ ($A_i$) such that the vector $\bfa_v$ defined as
    $$\bfa_v\cdot\bfG_{2} = \bfc_v^{(2)}\hspace{1cm}
      (\bfa_v\cdot\bfG_{i} = \bfc_v^{(i)})$$
has symbol $b$ in coordinate $\{v,w\}$. This quantity is passed on
through edge $\{v,w\}$ to the right decoder at vertex $w$.

In step two of stage $2$ (stage $i$), the right decoder at $w\in V_2$ ($w\in
V_i$) writes on its edge set $E(w)$
the codeword $\hat\bfb_2=(\hat\bfb_{2,j})_{j\in E(w)}$ 
($\hat\bfb_i=(\hat\bfb_{i,j})_{j\in E(w)}$)
of $B_2$ ($B_i$) that minimizes
$\sum_{j\in E(w)}d_j(\hat\bfb_{2,j})$ ($\sum_{j\in E(w)}d_j(\hat\bfb_{i,j})$).

In step three of stage $2$ (stage $i$), the decoder reverts to the basic
decoding scheme applied to the basic bipartite-graph code
$C( (V_0\cup V_2,E_2); A_{2,aux}, B_2 )$ 
(resp., $C( (V_0\cup V_i,E_i); A_{i,aux}, B_i )$).
   Call $\hat \bfa_v^{2}$ ( $\hat\bfa_v^{i}$) the vector recovered at vertex
   $v$ at the end of the basic decoding procedure.
For every $v\in V_0$ we then compute
        $$\bfy_{3,v}=\bfy_{2,v}+\hat\bfa_v^{2}\bfG_{2}
          \hspace{1cm}(\text{resp., }\bfy_{i+1,v}=\bfy_{i,v}+\hat\bfa_v^{i}\bfG_{i})$$
and name $\bfy_{3}$ ($\bfy_{i+1}$) the vector formed by the
$\bfy_{3,v}$ ($\bfy_{i+1,v}$), $v\in V_0$.

\medskip

The intuition behind the code construction and its decoding is as follows. 
In the $i$th
stage of decoding we would like to perform iterations of expander 
decoding for some expander code relying on the coordinates of the 
vector output by the $(i-1)$th stage on the coordinates of $E_i.$
This forms a difference with serial concatenations: there in 
the $i$th decoding stage we did not need to operate separately on 
the subset of coordinates of the code $A_1$; for expander codes
this is in the core of error correction. Note an analogy of (\ref{eq:coset2})
with (\ref{eq:coset}): every vector 
$\bfy_{1,v}$ can be written as 
$\bfy_{1,v}=\bfc_{1,v}+\bfe_v$ where $\bfc_{1,v}\in A_1$
is the code vector transmitted at the vertex $v$ and $\bfe_v$ is
the error vector added by the channel. Both procedures (\ref{eq:coset2})
and (\ref{eq:coset}) serve the goal of revealing a code vector 
$\bfy_{2,v}=\bfc_{2,v}+\bfe_v$, which is in the same relation to the code
$A_2$ as $\bfy_{1,v}$ is to $A_1.$

\subsection{Performance} The error probability of decoding for multilevel
codes on bipartite graphs is estimated analogously to the serial case.
In particular, the probability that the 
algorithm described in the previous section 
will result in a decoding error in the
$i$th stage, $i=1,\dots,m$ is estimated in Theorem \ref{thm:sl}.
Choosing the component codes as in the one-level parallel construction,
we conclude that the exponent of this probability approaches
(\ref{eq:fex}) as $n\to\infty.$ We will choose the code $A$ of sufficiently
large length $\Delta$ (independent of $n$) so that every code in the 
tower of codes (\ref{eq:subcodes2}) has error exponent of max-likelihood
decoding approaching the random coding exponent $E_0.$ 

The complexity of the whole procedure is essentially $m$ times that of
the single-level case.
To estimate the probability of decoding error $\cD$, let $\cD_i$ denote the
event that decoding is incorrect in the $i$th stage of the multistage
procedure. Then
    \begin{align*}
      P(\cD)&=P(\cD_m\cup\dots\cup\cD_1)\\
       &=
            P(\cD_m (\overline{\cD_{m-1}\cup\dots\cup
       \cD_{1}})) + P(\cD_m\,(\cD_{m-1}\cup\dots\cup
       \cD_{1})\,)\\
       &\le P(\cD_m | \;\overline{\cD_{m-1}\cup\dots\cup
       \cD_{1}})+ P(\cD_{m-1}\cup\dots\cup \cD_{1})\\
       &\le P(\cD_m |\; \overline{\cD_{m-1}\cup\dots\cup
       \cD_{1}})+\dots+ P(\cD_2|\overline \cD_1)+P(\cD_1).
    \end{align*}
As remarked above, Theorem \ref{thm:sl} implies that for all $i=1,\dots,m,$
    $$
      P(\cD_i|\;\overline{\cD_{i-1}\cap\dots\cap\cD_1})\le
         2^{-n\Delta t(E_0(R_{0,i})(1-R_{1,i})-\epsilon)}
    $$
(for $i=1$ the condition is empty). Choosing $\Delta_i=R_0\Delta/m$ for
all $i=1,\dots, m$ and repeating the argument that led
to Proposition \ref{prop:m-exp}, we obtain the main result of this paper.
\begin{theorem} Suppose that the $m$-level bipartite-graph codes are
used on a BSC$(p)$. For any rate $R<1-h(p)$ and any
$\epsilon>0$ there exists a family of
graphs $G$ of sufficiently large degrees $\Delta,\Delta_1,\dots,\Delta_m$ 
and a choice of component codes  such that the multistage
decoding algorithm of Sect. {\rm\ref{sect:decoding}}
has the error exponent $E^{(m)}(R,p)-\epsilon,$ where $E^{(m)}(R,p)$
is the error exponent of serially concatenated codes {\rm(\ref{eq:m-exp})}.
For large $m$ the error exponent approaches the Blokh-Zyablov bound 
{\rm (\ref{eq:inf-exp})}. The proportion of errors corrected by the 
algorithm approaches $(1/2)\delta_{\text{\rm BZ}}.$
\end{theorem}

\renewcommand\baselinestretch{0.9}
{\footnotesize
\providecommand{\bysame}{\leavevmode\hbox to3em{\hrulefill}\thinspace}

}
\end{document}